\documentclass[aps,prl,twocolumn,amsmath,amssymb]{revtex4}
\usepackage{bm}
\usepackage{graphicx}

\DeclareMathOperator{\sgn}{sgn}

\DeclareMathOperator{\Rea}{\mathrm{Re}}
\DeclareMathOperator{\Sp}{\mathrm{Sp}}

\newcommand{\beginsupplement} {
    \setcounter{table}{0}
    \renewcommand{\thetable}{S\arabic{table}}
    \setcounter{figure}{0}
    \renewcommand{\thefigure}{S\arabic{figure}}
    \setcounter{equation}{0}
    \renewcommand{\theequation}{S\arabic{equation}}
}

\begin{document}

\title{Phase-sensitive thermoelectricity and long-range Josephson effect supported by thermal gradient}
\author{Mikhail S. Kalenkov}
\affiliation{I.E. Tamm Department of Theoretical Physics, P.N. Lebedev Physical Institute, 119991 Moscow, Russia}
\author{Pavel E. Dolgirev}
\affiliation{Department of Physics, Harvard University, Cambridge Massachusetts 02138, USA}
\author{Andrei D. Zaikin}
\affiliation{Institut f{\"u}r Nanotechnologie, Karlsruher Institut f{\"u}r Technologie (KIT), 76021 Karlsruhe, Germany}
\affiliation{National Research University Higher School of Economics, 101000 Moscow, Russia}

\date{\today}
\begin{abstract}
We demonstrate that thermoelectric signal as well as dc Josephson current may be severely enhanced in multi-terminal superconducting hybrid nanostructures  exposed to a temperature gradient. At temperatures $T$ strongly exceeding the Thouless energy of our device both the supercurrent and the thermo-induced voltage are dominated by the contribution from non-equilibrium low energy quasiparticles and are predicted to decay slowly (algebraically rather than exponentially) with increasing $T$. We also predict a non-trivial current-phase relation and a transition to a $\pi$-junction state controlled by both the temperature gradient and the system topology. All these features are simultaneously observable in the same experiment.

\end{abstract}
\maketitle

Superconducting hybrid structures exposed to a temperature gradient acquire a variety of intriguing properties. One of them is the thermoelectric effect \cite{Ginzburg} implying the presence of thermo-induced electric currents and/or voltages inside the sample.
At low temperature these thermoelectric signals are {\it phase-coherent} which results in their periodic dependence on the phase of a superconducting condensate. Thermoelectricity gives rise to diverse applications ranging from thermometry and refrigeration
\cite{Giazotto} to phase-coherent caloritronics \cite{FG} paving the way to an emerging field of thermal logic \cite{Li} operating with information in the form of energy.

Superconducting circuits appropriate for such applications may involve superconducting-normal-superconducting (SNS) junctions of different geometry. In such structures low temperature electron transport is strongly influenced by the proximity effect implying penetration of superconducting correlations deep into normal metal. As a result, macroscopic quantum coherence is established across the whole structure thus supporting the Josephson current $I_J$ between superconducting terminals. 

In equilibrium, the magnitude of this effect essentially depends on the relation between temperature $T$ and an effective Thouless energy $ E_{\mathrm{Th}}$ of an SNS device. As soon as $T$ strongly exceeds $ E_{\mathrm{Th}}$ the supercurrent reduces exponentially $I_J \propto e^{-\sqrt{2 \pi T/E_{\mathrm{Th}}}}$ \cite{BWBSZ1999,GKI} and, hence, long-range phase coherence gets effectively suppressed at such values of $T$. A similar conclusion concerning the magnitude of the thermoelectric voltage signal $V_T$ could be extracted from a number of previous 
theoretical studies \cite{SV,VH,KZ17}.

In this Letter we will demonstrate that by exposing the system to a temperature gradient one can effectively support long-range phase coherence
at temperatures strongly exceeding the Thouless energy $E_{\mathrm{Th}}$ where the equilibrium supercurrent becomes negligible. 

Consider a long SNS junction with normal state resistance $R_n$ and two extra normal terminals attached to the central N-wire as shown in Fig. 1. Provided these normal terminals are maintained at different temperatures $T_1$ and $T_2$ the electron distribution function inside the junction is driven out of equilibrium. Below we are going to demonstrate that in the limit $T_{1,2}\gg E_{\mathrm{Th}}$ the Josephson critical current $I_C$ -- up to some geometry factors -- takes the form 
\begin{equation}
I_C \sim  E^2_{\mathrm{Th}}|1/T_1 - 1/T_2|/(eR_n),
\label{1}
\end{equation} 
thus being a lot bigger that the equilibrium current $I_J$ at any of the two temperatures $T_1$ or $T_2$. In addition, in this regime the system is described by a non-sinusoidal current-phase relation (CPR) and may exhibit a pronounced $\pi$-junction-like behavior.

Furthermore, below we will show that -- depending on its topology -- the system can develop a large phase-coherent thermoelectric voltage
signal that {\it does not} decay exponentially even if temperature increases above  $E_{\mathrm{Th}}$. Remarkably, at $T_{1,2}\gg E_{\mathrm{Th}}$  the magnitude of this signal $V_T$ turns out to have exactly the same temperature dependence as $I_C$, i.e.
\begin{equation}
V_T \sim I_CR_n.
\label{2}
\end{equation}
Both results (\ref{1}) and (\ref{2}) are due to the presence of non-equilibrium low energy quasiparticles suffering little dephasing while propagating across the system.

\begin{figure}
\begin{center}
\includegraphics[width=70mm]{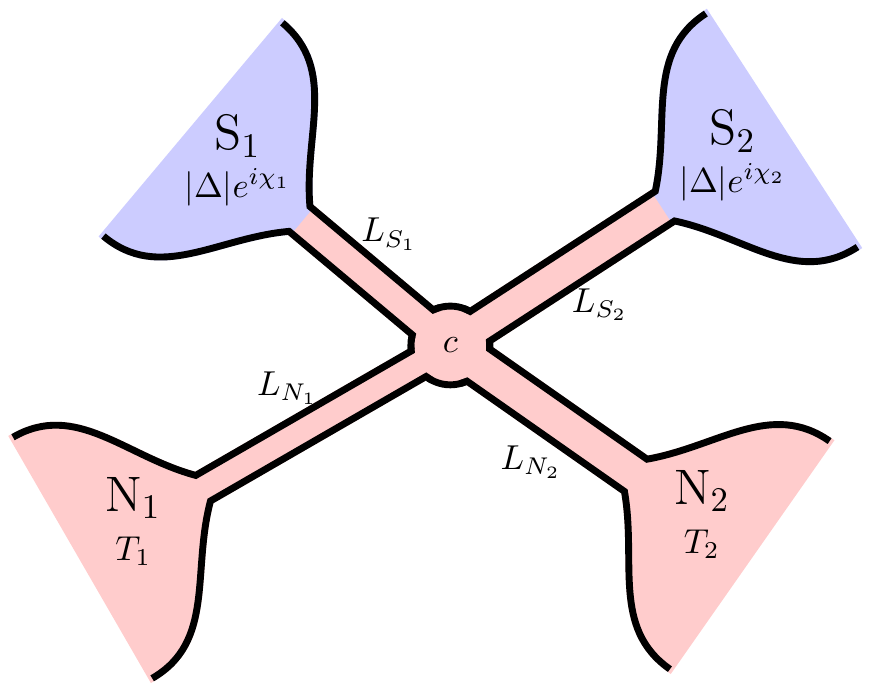}
\end{center}
\caption{X-junction structure under consideration.}
\label{ssXnn-fig}
\end{figure}

{\it The model and basic formalism.} We will consider the structure displayed in Fig. 1. It consists of two superconducting and two normal terminals interconnected by four normal metallic wires of lengths $L_{S_{1,2}}$, $L_{N_{1,2}}$ and cross sections ${\mathcal A}_{S_{1,2}}$, ${\mathcal A}_{N_{1,2}}$ respectively. For brevity in what follows we will denote this structure as {\it X-junction}.  The superconducting terminals are biased by the phase twist $\chi=\chi_1-\chi_2$ and the supercurrent $I_S(\chi)$ can flow between these terminals for nonzero 
$\chi$. The two normal terminals are disconnected from any external circuit and are maintained at different temperatures $T_1$ and $T_2$.

In order to proceed we will make use of the standard quasiclassical formalism of Usadel equations \cite{BWBSZ1999}
\begin{equation}
i D \nabla \left(\check G  \nabla \check G\right)=
\left[\hat \Omega\check 1, \check G \right], \quad \check G \check G=\check 1,
\end{equation}
which allow to evaluate $4\times4$ Green-Keldysh matrix functions $\check G=\left(\begin{smallmatrix}
\hat G^R & \hat G^K \\
0 & \hat G^A \\
\end{smallmatrix}\right)$ for our $X$-junction. Here $D$ stands for diffusion constant, $\hat G^{R,A}=\left(\begin{smallmatrix}
G^{R,A} & F^{R,A} \\
\tilde F^{R,A} & - G^{R,A}\\
\end{smallmatrix}\right)$ are retarded and advanced $2\times2$ Green function matrices in the Nambu space, $\hat \Omega=
\left(\begin{smallmatrix}
\varepsilon + eV & \Delta \\
- \Delta^* & -\varepsilon+ eV \\
\end{smallmatrix}\right)$,
where $\varepsilon$, $V$ and $\Delta$ denote respectively quasiparticle energy, electrostatic potential and superconducting order parameter. 
The Keldysh matrix has the form $\hat G^K = \hat G^R \hat h - \hat h \hat G^A$, where $\hat h$ is the matrix distribution function. The current density $\bm{j}$ is expressed by means of the standard relation
\begin{gather}
\bm{j} =  -\dfrac{\sigma}{8 e }
\int
d \varepsilon
\Sp (\hat \tau_3\check G \nabla \check G)^K,
\end{gather}
where $\sigma$ is the normal Drude conductivity and $\hat\tau_3$ is one of the Pauli matrices in the Nambu space.

It is convenient to decompose the matrix distribution function as
$\hat h = h^L + \hat \tau_3 h^T$. In the normal wires the functions $h^L$ and $h^T$ obey the diffusion-like equations
\begin{gather}
iD\nabla\left[ D^T\nabla h^T + {\mathcal Y}\nabla h^L + \bm{j}_{\varepsilon} h^L \right]=0,
\label{Tkin}
\\
iD\nabla\left[ D^L\nabla h^L - {\mathcal Y} \nabla h^T + \bm{j}_{\varepsilon} h^T \right]=0.
\label{Lkin}
\end{gather}
Here $D^{T/L}=\nu^2 \pm |F^R \pm F^A|^2/4$ define the two kinetic coefficients and $\nu = \Rea G^R$ is the local electron density of states.
The third kinetic coefficient ${\mathcal Y}=(|\tilde F^R|^2 |-F^R|^2)/4$ accounts for the presence of particle-hole asymmetry in our system and
\begin{equation}
\bm{j}_{\varepsilon}=\dfrac{1}{2}\Rea\left(F^R \nabla \tilde F^R - \tilde F^R \nabla F^R \right)
\end{equation}
defines the spectral current.

As usually, the above equations should be supplemented by proper boundary conditions at inter-metallic interfaces. Here we assume that all interfaces between the wires and the terminals are fully transparent and, hence, the Green functions are matched continuously at these interfaces. The same applies to the contact between the wires (point $c$ in Fig. 1). We also assume that all four normal wires are thin enough and long enough enabling one (a) to fully ignore their effect on the bulk terminals and (b)  to consider the effective 
Thouless energy of our device $E_{\mathrm{Th}}=D/L_S^2$ (with $L_S= L_{S_{1}}+L_{S_{2}}$) as the only relevant energy scale in our problem. This is appropriate provided $E_{\mathrm{Th}}  \ll |\Delta |$. The latter inequality -- combined with the condition $T_{1,2} \ll |\Delta |$ -- implies that our analysis can be restricted to subgap energies.

{\it Long-range phase coherent thermoelectricity.} 
Applying a thermal gradient to normal terminals N$_1$ and N$_2$ one induces thermoelectric voltages $V_1$ and $V_2$ at these terminals
\cite{SV,VH,KZ17,KPV,VP,DKZ2018,dolgirev2018topology}. These voltage signals are in general not small and depend periodically on the phase $\chi$, as it was repeatedly observed in experiments \cite{Venkat1,Petrashov1,Venkat2,Petrashov2}. Both these features are direct consequences of the particle-hole asymmetry generated by the mechanism of sequential Andreev reflection at two NS interfaces \cite{KZ17}. 

The quasiparticle distribution function inside the X-junction is recovered from the diffusion-like equations \eqref{Tkin}, \eqref{Lkin} combined with the observation that no electric current can flow into normal terminals N$_1$ and N$_2$. With this in mind we get 
\begin{equation}
h^{T/L}_{N_{1,2}}
=
\dfrac{1}{2}
\left[
\tanh \dfrac{\varepsilon + e V_{1,2}}{2 T_{1,2}}
\mp
\tanh \dfrac{\varepsilon - e V_{1,2}}{2 T_{1,2}}
\right]
\label{bou1}
\end{equation}
at the interfaces between the N-wire and the corresponding N-terminal, while at both at \textrm{SN} interfaces we have $h^T=0$. 

To begin with, we note that in partially symmetric X-junctions with (i) $L_{S_1} = L_{S_2}=L_S/2$ and (ii) ${\mathcal A}_{S_1} = {\mathcal A}_{S_2}$
the kinetic coefficient ${\mathcal Y}$ equals to zero in the crossing point $c$ and everywhere in the N-wires attached to normal terminals N$_1$ and N$_2$. In order prove this property one should bear in mind that ${\mathcal Y}$ is  an $odd$ function of the phase $\chi$. Interchanging the terminals S$_1 \leftrightarrow$ S$_2$ and inverting the phase sign $\chi \to -\chi$,
under the conditions (i) and (ii) we arrive at exactly the same X-junction as the initial one. Hence, in this case ${\mathcal Y}$ should also be an $even$ function of $\chi$ which is only possible if ${\mathcal Y} \equiv 0$.

\begin{figure}[t!]
\includegraphics[scale=1]{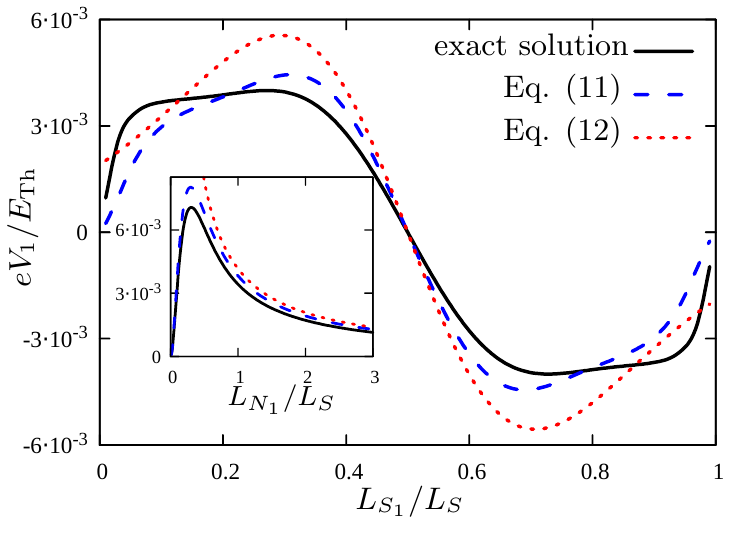}
\caption{Thermoelectric voltage $V_1$ at $\chi= \pi/2$, $T_1 = 20 E_{\mathrm{Th}} $ and  $T_2 = 30 E_{\mathrm{Th}} $ as a function of $L_{S_1}$ for $L_{N_1}=L_{N_2}=L_S$ and $\mathcal{A}_{S_1} = \mathcal{A}_{S_2} = \mathcal{A}_{N_1} = \mathcal{A}_{N_2}$. Inset: The same voltage as a function  of $L_{N_1}=L_{N_2}$ for $L_{S_1}=0.3L_S$.}
\label{V1-LS1-2-fig}
\end{figure}

Setting ${\mathcal Y}=0$ in Eqs. \eqref{Tkin}, \eqref{Lkin} in the wires connected to the N-terminals one may verify that  $h^T \equiv 0$ becomes a trivial solution of these equations everywhere in our system. Combining this solution with Eq. (\ref{bou1}) we observe that both voltages $V_{1,2}$ vanish identically in this case. Then the kinetic equations \eqref{Tkin}, \eqref{Lkin} reduce to $D^L \nabla h^L = C_1$ and $h^L  = C_2$ (with $C_1$ and $C_2$ being constants) in the N-wires connected respectively to N- and to S-terminals. Resolving these equations we recover the distribution function $h^L$ inside the wires attached to the superconducting terminals:
\begin{equation}
h^L = r^L_{N_2}h^L_{N_1}+r^L_{N_1}h^L_{N_2},
\label{hc}
\end{equation}
where $r^L_{N_i}=R^L_{N_i}/(R^L_{N_1} + R^L_{N_2})$ and 
\begin{equation}
R^L_{N_i} = 
\dfrac{1}{\mathcal{A}_{N_i}\sigma} \int\limits_{L_{N_i}} \dfrac{dx}{D^L}, \quad i=1,2
\end{equation}
are spectral resistances of the N-wires attached to the normal terminals N$_1$ and N$_2$.

The above simple analysis demonstrates that no thermoelectric effect may occur in our X-junction provided the kinetic coefficient ${\mathcal Y}$ vanishes in the N-wires attached to the normal terminals. We now lift the conditions (i), (ii) and evaluate the thermoelectric voltages $V_1$ and $V_2$. 

\begin{figure}[t!]
\includegraphics[scale=1]{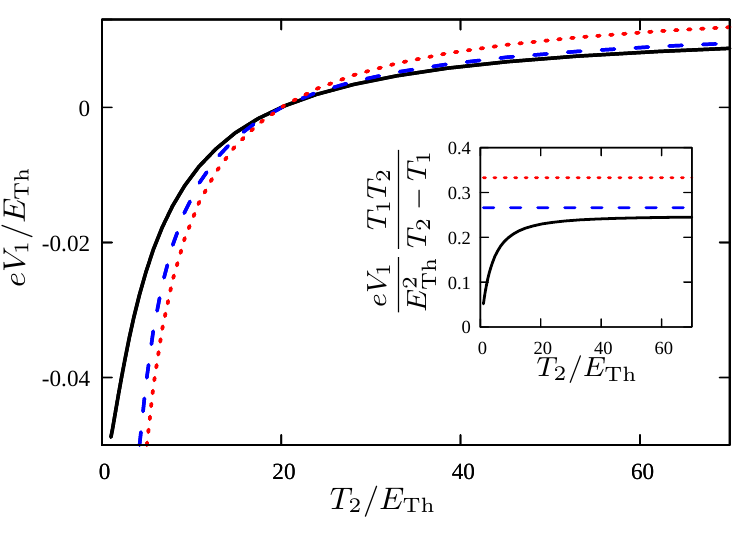}
\caption{Thermoelectric voltage $V_1$ as a function of temperature $T_2$. The notations and the values  of $\chi$, $T_1 $, $L_{S_1}$ are the same as in Fig. 2 and  $L_{N_1}=L_S/2$, $L_{N_2}=L_S$.}
\label{V1-T-ver2-fig}
\end{figure}

The corresponding derivation is outlined in Supplemental Materials, here we only quote the final result. Assuming that both temperatures strongly exceed the Thouless energy $T_{1,2} \gg E_{\mathrm{Th}}$, for the thermoelectric voltage induced at the terminal N$_1$ we obtain
\begin{gather}
e V_{1} = \dfrac{r_{N_1}}{4}
\left(
\dfrac{1}{T_2} - \dfrac{1}{T_1}
\right)
\int \varepsilon d \varepsilon \int_{L_{N_1}}{\mathcal Y}
\dfrac{dx }{L_{N_1}},
\label{thV}
\end{gather}
where $r_{N_i}=R_{N_i}/(R_{N_1} + R_{N_2})$ and $R_{N_i} = L_{N_i} / (\mathcal{A}_{N_i}\sigma)$ ($i=1,2$) are normal 
state resistances of the wires attached to normal reservoirs. The function $\mathcal{Y}$ in Eq. (\ref{thV}) can be evaluated numerically or
estimated analytically extrapolating the results derived in the limit $|\varepsilon| \gg E_{\mathrm{Th}}$ to lower energies. The latter procedure
allows to perform the integrals in Eq. \eqref{thV} and get
\begin{multline}
e V_{1} \approx
\dfrac{\gamma\varkappa^2 r_{N_1}(L_{S_1} - L_{S_2})E_{\mathrm{Th}}^2}{L_{N_1}(3 + 2\sqrt{2})}
\left(\dfrac{1}{T_2} - \dfrac{1}{T_1}\right)\sin \chi,
\label{V1theor}
\end{multline}
where $\gamma=( L_{S}^2  + 2L_{S_1} L_{S_2}) L_{S}^4/(L_{S_1}^2 + L_{S_2}^2)^3$ and $\varkappa = 4 \sqrt{\mathcal{A}_{S_1}\mathcal{A}_{S_2}}/(\mathcal{A}_{S_1}+\mathcal{A}_{S_2}+ \mathcal{A}_{N_1} + \mathcal{A}_{N_2})$ are dimensionless geometric factors.

Equations (\ref{thV}), (\ref{V1theor}) represent the first key result of our present work. The periodic dependence of the thermoelectric signal (\ref{V1theor}) on the phase $\chi$ demonstrates that long-range phase coherence in our X-junction is well maintained even at high enough temperatures $T_{1,2} \gg E_{\mathrm{Th}}$. It is also remarkable that under this condition the amplitude of the thermo-induced voltage $V_1$ (\ref{V1theor}) decreases with increasing temperature only as a power-law, i.e. much slower than it was previously reported elsewhere \cite{SV,VH,KZ17}. 

The thermoelectric voltage $V_2$ induced at the second normal terminal N$_2$ can be obtained from the above Eqs. (\ref{thV}), (\ref{V1theor}) by interchanging the indices $1 \leftrightarrow 2$. In symmetric structures with $L_{N_1} = L_{N_2}$ and ${\mathcal A}_{N_1} = {\mathcal A}_{N_2}$ one readily finds $V_2=-V_1$.

In addition to the above analysis we resolved the Usadel equations numerically and evaluated the thermoelectric voltages $V_{1,2}$ employing no approximations. Our numerically exact results for $V_1$ are displayed in Figs. 2 and 3 (solid lines) together with Eq.  (\ref{thV}) (where ${\mathcal Y}$ was evaluated numerically) and Eq. (\ref{V1theor}) indicated respectively by long and short dashed lines.

{\it Long-range Josephson effect.} We now turn to dc Josephson effect in the presence of a temperature gradient.
For simplicity in what follows we again impose the symmetry conditions (i), (ii) and denote ${\mathcal A}_{S_{1,2}} = {\mathcal A}_{S}$. As we demonstrated above, in this particular case no electron-hole asymmetry is generated and, hence, no thermoelectric effect occurs, i.e. $V_{1,2}=0$. Furthermore, the distribution function $h^T$ equals to zero, while the function $h^L$ inside the wires is defined by Eq. \eqref{hc}. 

Let us introduce the function $W(\varepsilon)=r^L_{N_2} r_{N_1} - r^L_{N_1} r_{N_2}$ and identically rewrite the latter equation in the form
\begin{equation}
h^L = r_{N_2}h^L_{N_1} + r_{N_1}h^L_{N_2}+
W(\varepsilon) (h^L_{N_1} - h^L_{N_2}).
\label{hcW}
\end{equation} 
The first two terms in the right-hand side of Eq. (\ref{hcW}) represent a superposition of the equilibrium distribution functions $h^L_{N_1}$ and $h^L_{N_2}$ with energy independent prefactors, while the last term is essentially non-equilibrium in nature. The function $W(\varepsilon)$ vanishes identically in structures with $L_{N_1} = L_{N_2} $, otherwise it remains nonzero at low enough energies and decays exponentially provided $|\varepsilon |$ exceeds the Thouless energy of our device $E_{\mathrm{Th}}$.

With the aid of Eqs. (\ref{hcW}) we immediately recover the expression for the supercurrent $I_S$ flowing between the superconducting terminals S$_1$ and S$_2$ across the normal wire of length $L_S$. We obtain
\begin{equation}
I_S = r_{N_2} I_J(T_1,\chi) + r_{N_1} I_J(T_2,\chi)+I_S^{\rm ne}(T_1,T_2,\chi),
\label{IS}
\end{equation}
where 
\begin{equation}
I_J(T,\chi)= - \dfrac{\sigma \mathcal{A}_S}{2e}\int j_{\varepsilon} \tanh \dfrac{\varepsilon}{2T}
d \varepsilon
\label{IJ}
\end{equation}
is the equilibrium Josephson current and 
\begin{equation}
I_S^{\rm ne}=\dfrac{\sigma\mathcal{A}_S }{2e}
\int
j_{\varepsilon}
W ( \varepsilon)
\left(
\tanh \dfrac{\varepsilon}{2T_2} - \tanh \dfrac{\varepsilon}{2T_1}
\right)
d \varepsilon .
\label{ISne}
\end{equation}

Equations \eqref{IS}-\eqref{ISne} define the second key result of this work. It demonstrates that provided our X-junction is biased by a temperature gradient the supercurrent $I_S$  consists of two different contributions. The first one is a weighted sum of equilibrium Josephson currents $I_J$ (\ref{IJ}) evaluated at temperatures $T_1$ and $T_2$ and the second one $I_S^{\rm ne}$ \eqref{ISne} accounts specifically for non-equilibrium effects. 

It is easy to verify that provided at least one of the two temperatures remains below the Thouless energy $E_{\mathrm{Th}}$ the current $I_S$ \eqref{IS} is dominated by the first (quasi-equilibrium) contribution, while the non-equilibrium one (\ref{ISne}) can be safely neglected. On the other hand, at $T_{1,2} \gg E_{\mathrm{Th}}$ the equilibrium contribution to $I_S$ gets exponentially suppressed as (cf. \cite{ZZh,GreKa}):
\begin{multline}
I_J= \dfrac{16 \varkappa}{ 3 + 2\sqrt{2}}\dfrac{E_{\mathrm{Th}}}{eR_n}\left(\dfrac{2 \pi T}{E_{\mathrm{Th}}}\right)^{3/2}
e^{-\sqrt{2 \pi T /E_{\mathrm{Th}}}} \sin\chi ,
\label{IJeq}
\end{multline}
where $R_n=L_S/(\mathcal{A}_S\sigma )$ and the prefactor $\varkappa$ is taken at $\mathcal{A}_{S_{1,2}} =\mathcal{A}_S$. Thus, at $T_{1,2} \gg E_{\mathrm{Th}}$ the supercurrent can already be dominated by the non-equilibrium term  $I_S^{\rm ne}$. Evaluating the energy integral in Eq. (\ref{ISne}) we obtain 
\begin{multline}
I_S^{\rm ne} \simeq
0.21\varkappa^3 r_{N_1} r_{N_2}\dfrac{ E^2_{\mathrm{Th}}}{eR_n}\left(\dfrac{1}{T_1} - \dfrac{1}{T_2} \right)\\\times\left(\dfrac{L_S}{L_{N_2}}-\dfrac{L_S}{L_{N_1}}\right) 
\sin \chi \cos^2(\chi/2).
\label{Wapprox}
\end{multline}

This result is remarkable in several important aspects. First of all, we observe that at temperatures strongly exceeding the Thouless energy the  supercurrent $I_S\simeq I_S^{\rm ne}$ decays with increasing min$(T_1,T_2)$ only as a power law unlike the equilibrium Josephson current 
in long SNS junctions which is known to decay exponentially. This behavior is clearly due to driving the electron distribution function  $h^L$ out of equilibrium by applying a temperature gradient. Keeping $T_1$ fixed, we observe that the supercurrent magnitude  $grows$ with $T_2$ (cf. also Figs. 4 and 5) strongly exceeding the equilibrium value $I_J$ (\ref{IJeq}) at any of the two temperatures $T_1$ or $T_2$. Hence, we predict {\it strong supercurrent stimulation by a temperature gradient}.

\begin{figure}
\includegraphics[width=80mm]{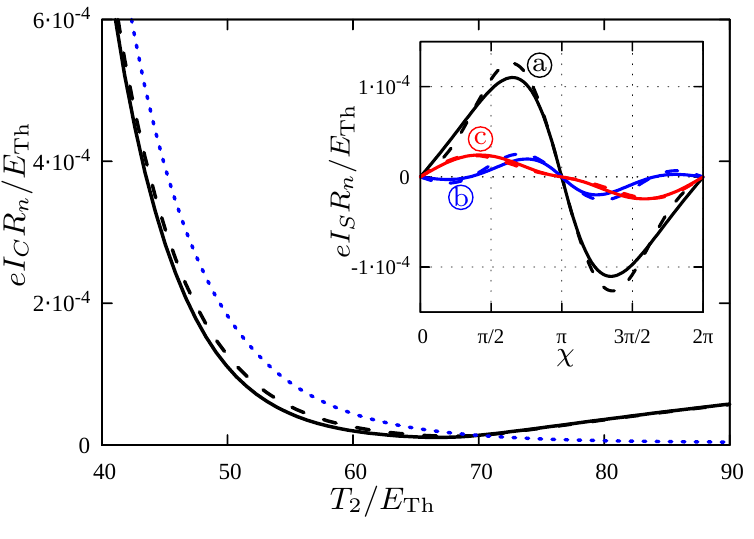}
\caption{Josephson critical current $I_C\equiv {\rm max}|I_S|$ as a function of $T_2$. Inset: CPR evaluated at $T_2= 50 E_{\mathrm{Th}}$ (a),  $60 E_{\mathrm{Th}}$ (b) and $75 E_{\mathrm{Th}}$ (c). Solid lines correspond to the exact numerical solution, dashed lines indicate the result (\ref{IS}) combined with (\ref{IJeq}) and (\ref{Wapprox}), dotted line is the quasi-equilibrium contribution $r_{N_2} I_J(T_1,\pi/2) + r_{N_1} I_J(T_2,\pi/2)$ to $I_S$. The parameters are: $T_1 = 70 E_{\mathrm{Th}}$, $L_{S_{1,2}}=L_S/2$, $L_{N_1}=3 L_S$, $L_{N_2}=L_S$ and $\mathcal{A}_{S_1} = \mathcal{A}_{S_2} = \mathcal{A}_{N_1} = \mathcal{A}_{N_2}$. }
\label{Ic-T-chi-3-1-fig}
\end{figure}

Another interesting feature of the result (\ref{Wapprox}) is the non-sinusoidal CPR that persists at temperatures well above
$E_{\mathrm{Th}}$. For comparison, the dependence of the equilibrium Josephson current on the phase $\chi$ in SNS junctions remains 
non-sinusoidal only at $T \lesssim E_{\mathrm{Th}}$ and reduces to $I_J \propto \sin \chi$ at higher temperatures.

In addition, we observe that the sign of the supercurrent in Eq. (\ref{Wapprox}) is controlled by those of both length and temperature differences, $L_{N_1} - L_{N_2}$ and $T_1 - T_2$. For instance, by choosing $L_{N_1} < L_{N_2}$ and $T_1 <T_2$ we arrive at a pronounced $\pi$-junction-like behavior, see also Fig. 5.

\begin{figure}
\includegraphics[width=80mm]{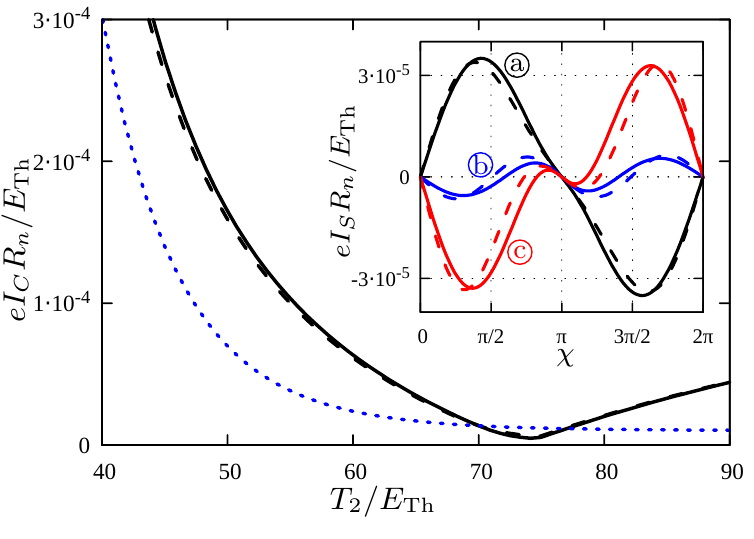}
\caption{The same as in Fig. 4. The parameters are the same except $L_{N_1}=L_S$, $L_{N_2}=3 L_S$. Temperature values in the inset are $T_2= 65 E_{\mathrm{Th}}$ (a),  $75 E_{\mathrm{Th}}$ (b) and $85 E_{\mathrm{Th}}$ (c).}
\label{}
\end{figure}

Previously switching to the $\pi$-junction state in a configuration similar to ours was realized by applying at external voltage bias $V$ to normal terminals \cite{Volkov,WSZ,Yip,Teun}. In this case the electron distribution function is also driven out of equilibrium, however, unlike here,
the magnitude of the supercurrent remains exponentially small for $eV,T \gg E_{\mathrm{Th}}$ \cite{WSZ}. On the other hand, by creating non-equilibrium conditions with the aid of an external rf-signal it is possible to efficiently stimulate the supercurrent in long SNS junctions \cite{Aslamazov82,Zaikin83}, however, no $\pi$-junction behavior could be obtained in this way. In contrast to the above examples, exposing the X-junction to a temperature gradient makes both non-trivial features -- supercurrent stimulation and $\pi$-junction states -- simultaneously observable in the same experiment.

The supercurrent $I_S$ was also evaluated numerically without employing any approximations. The corresponding results are displayed in Figs. 4 and 5 together with Eq. (\ref{IS}) combined with  Eqs. (\ref{IJeq}) and (\ref{Wapprox}). In Fig. 4 the parameters are chosen such that the non-equilibrium term $I_S^{\rm ne}$ remains negative for $T_2<T_1$ and the $\pi$-junction states may only exist in a tiny interval of $T_2$ below $T_1$. In contrast, in Fig. 5 the $\pi$-junction behavior is realized practically at any  $T_2>T_1$ (cf. curves (b) and (c) in the inset) since the term $I_S^{\rm ne}$ takes negative values at such temperatures. 

Note that -- in contrast to the standard situation -- here the transition between 0- and $\pi$-junction states does $not$ correspond to vanishing
Josephson critical current $I_C\equiv {\rm max}|I_S|$ because of a non-sinusoidal form of CPR (\ref{Wapprox}). The value $I_C$ may be achieved either at $\chi < \pi/2$ or at $\chi >\pi/2$ depending on whether the maximum or the minimum of $I_S^{\rm ne}$ (\ref{Wapprox}) is reached at $\chi =\pi/3$. Furthermore, the competition between the terms $\propto I_J$ and  $I_S^{\rm ne}$ may also cause extra maximum and minimum of the dependence $I_S(\chi)$ (cf. curve (b) in Fig. 4 and curves (b), (c) in Fig. 5), since in a narrow vicinity of $\chi=\pi$ the contribution containing $I_J \propto (\pi -\chi)$
always dominates over the non-equilibrium one $I_S^{\rm ne} \propto (\pi - \chi)^3$.

In summary, we have demonstrated that clear manifestations of long-range phase coherence may persist up to much higher temperatures as compared to the Thouless energy $E_{\mathrm{Th}}$ provided our X-junction is exposed to a temperature gradient. In particular, at $T_{1,2}\gg E_{\mathrm{Th}}$
both the Josephson critical current $I_C$ and the magnitude of the phase-coherent voltage signal $V_T={\rm max}V_{1,2}$ exhibit exactly the same algebraic dependence on $T_1$ and $T_2$, cf. Eqs. (\ref{1}) and (\ref{2}). In both cases long-range phase coherence is maintained due to
non-equilibrium quasiparticles with energies below $E_{\mathrm{Th}}$ propagating across the system without any significant phase relaxation \cite{FN}. Our results indicate that quantum properties of X-junctions and similar hybrid structures can be efficiently controlled and manipulated with the aid of both superconducting phase and temperature gradient.

Two of us (M.S.K. and A.D.Z.) acknowledge partial support by RFBR grant No. 18-02-00586.

\beginsupplement
\section{Supplemental Materials}

\subsection{I. High energy expansion for the Green functions}

At energies exceeding the relevant Thouless energy of our device it is possible to recover approximate analytic expressions for retarded and advanced Green functions. Inside normal metallic wires at relatively short distances $x \ll L_S$ away from the interface with one of the superconducting terminal one can safely disregard the effect of another such terminal.  Then the anomalous Green function can be written in the form
\begin{equation}
F^R = -i \dfrac{4 y (1- y^2)}{(1+y^2)^2} e^{i\chi_i},
\label{FRSN}
\end{equation}
where $\chi_i$ is the phase of the order parameter in the nearest superconducting terminal and 
\begin{equation}
y = a_S e^{-\sqrt{-2i\varepsilon/D}x},
\,
a_S (\varepsilon) = 
\tan\left[\dfrac{1}{4}\arcsin \dfrac{|\Delta|}{\sqrt{|\Delta|^2 - \varepsilon^2}}\right].
\end{equation}
At distances from both S-terminals exceeding $\sqrt{D/|\varepsilon|}$ the anomalous Green functions take exponentially small values 
enabling one to linearize the Usadel equation, i.e.
\begin{equation}
(F^R)'' + \dfrac{2 i \varepsilon}{D}F^R=0.
\label{FRlinear}
\end{equation}
Matching the solution of Eq. (S3) with the asymptotics (S1) in the vicinity of SN interfaces one recovers the anomalous Green function at the crossing point $c$:
\begin{multline}
F_c =
-\dfrac{8i \mathcal{A}_{S_1} a_{S_1} e^{-\sqrt{-2i\varepsilon/D}L_{S_1}} e^{i\chi_1}}{
\mathcal{A}_{S_1} + \mathcal{A}_{S_2} + \mathcal{A}_{N_1} + \mathcal{A}_{N_2}
}
-\\-
\dfrac{8i \mathcal{A}_{S_2} a_{S_2} e^{-\sqrt{-2i\varepsilon/D}L_{S_2}} e^{i\chi_2}}{
\mathcal{A}_{S_1} + \mathcal{A}_{S_2} + \mathcal{A}_{N_1} + \mathcal{A}_{N_2}
}
\label{Fc}
\end{multline}
and in the normal wires connected to the normal terminals
\begin{equation}
F^R = F_c e^{-\sqrt{-2i\varepsilon/D}x},
\end{equation}
where $x$ is the distance from the crossing point.

\subsection{II. Long-range thermoelectric effect}
Kinetic equations (5), (6) in the quasi-one-dimensional geometry can be rewritten as
\begin{gather}
D^T ( h^T)' + \mathcal{Y} (h^L)' + j_{\varepsilon} h^L =-eI^T/(\sigma \mathcal{A}),
\label{Tkin2}
\\
D^L (h^L)' - \mathcal{Y} (h^T)' + j_{\varepsilon} h^T =-eI^L/(\sigma \mathcal{A}),
\label{Lkin2}
\end{gather}
where $\mathcal{A}$ is the wire cross section, $I^T$ and $I^L$ are the spectral currents. For the sake of definiteness we choose the current to be positive provided it flows from the corresponding terminal to the crossing point $c$. Equations (S6), (S7) establish linear relations between the distribution functions $h^T$ and $h^L$ at the ends of the wire segments. For the wires connected to the normal terminals we get
\begin{equation}
\begin{pmatrix}
G^T_{N_i} & G^{\mathcal{Y}}_{N_i} \\
- G^{\mathcal{Y}}_{N_i} & G^L_{N_i}
\end{pmatrix}
\begin{pmatrix}
h^T_c - h^T_{N_{i}}
\\
h^L_c - h^L_{N_{i}}
\end{pmatrix}
=
\begin{pmatrix}
I^T_{N_i} \\ I^L_{N_i}
\end{pmatrix},
\quad i =1,2,
\end{equation}
where $G^T_{N_i}$, $G^L_{N_i}$ and $G^{\mathcal{Y}}_{N_i}$ are spectral conductances defined as
\begin{equation}
\begin{pmatrix}
G^T_{N_i} & G^{\mathcal{Y}}_{N_i} \\
- G^{\mathcal{Y}}_{N_i} & G^L_{N_i}
\end{pmatrix}
=
\left[
\int_{L_{N_i}}\begin{pmatrix}
D^T & \mathcal{Y} \\
- \mathcal{Y} & D^L
\end{pmatrix}^{-1}
\dfrac{dx}{ \sigma \mathcal{A}_{N_i} }
\right]^{-1}.
\end{equation}
These conductances $G^T_{N_i}$, $G^L_{N_i}$ and $G^{\mathcal{Y}}_{N_i}$ exhibit a nontrivial energy dependence in the vicinity of the Thouless energy. In the high energy limit $G^{\mathcal{Y}}_{N_i}$ tends to zero and $G^T_{N_i}$, $G^L_{N_i}$ just reduce to normal state wire conductances $\sigma \mathcal{A}_{N_i}/ L_{N_i}$. 

In the wires connected to superconducting terminals and at subgap energies the spectral currents $I^L_{S_i}$ vanish identically. This observation helps to simplify the relations between the distribution functions, which now read
\begin{gather}
G^T_{S_1} h_c^T + \mathcal{A}_{S_1} \sigma j_{1\varepsilon} h_c^L = -e I^T_{S_1},
\\
G^T_{S_2} h_c^T - \mathcal{A}_{S_2} \sigma j_{2\varepsilon} h_c^L = -e I^T_{S_2}.
\end{gather}
Here we also made use of the fact that  the distribution function $h^T$ equals to zero at both SN interfaces at subgap energies. 

In general the spectral conductances $G^T_{S_{1,2}}$ depend on the kinetic coefficients $D^{T,L}$ and $\mathcal{Y}$ in a complicated manner. These conductances demonstrate a nontrivial energy dependence at energies below the Thouless one and tend to normal state wire conductances $\sigma \mathcal{A}_{S_{1,2}}/ L_{S_{1,2}}$ in the high energy limit. The spectral current conservation conditions at the crossing point take the form
\begin{gather}
I^T_{S_1} + I^T_{S_2} + I^T_{N_1} + I^T_{N_2} =0,
\\
I^L_{N_1} + I^L_{N_2} =0, 
\quad
\mathcal{A}_{S_1} j_{1\varepsilon} = \mathcal{A}_{S_2} j_{2\varepsilon}.
\end{gather}

Consider first a symmetric X-junction with $L_{N_1} = L_{N_2} = L_{N}$ and ${\mathcal A}_{N_1} = {\mathcal A}_{N_2} = {\mathcal A}_{N}$. In this case the distribution function $h^T_c$ at the crossing point reads
\begin{equation}
h^T_c = \dfrac{
[G_N^T G_N^L + (G_N^{\mathcal{Y}})^2] (h_{N_1}^T + h_{N_2}^T)
}{
(2 G_N^T + G^T_{S_1} + G^T_{S_2}) G_N^L + 2 (G_N^{\mathcal{Y}})^2
}.
\label{hcS}
\end{equation}
With the aid of the current conservation conditions we get
\begin{equation}
\int (G^T_{S_1} + G^T_{S_2}) h^T_c d \varepsilon = 0.
\label{cond0}
\end{equation}
Combining Eqs. (S14) and (S15) one readily finds
\begin{gather}
\int Q(\varepsilon)  (h_{N_1}^T + h_{N_2}^T)
d \varepsilon = 0,
\label{cond1}
\\
Q(\varepsilon)= \dfrac{
[G_N^T G_N^L + (G_N^{\mathcal{Y}})^2] (G^T_{S_1} + G^T_{S_2})
}{
(2 G_N^T + G^T_{S_1} + G^T_{S_2}) G_N^L + 2 (G_N^{\mathcal{Y}})^2
}.
\end{gather}
On the other hand, since no current can flow into or out of the normal terminals, we may write
\begin{equation}
\int G^T_{N} (h_{N_1}^T - h_{N_2}^T) d \varepsilon 
+
\int G^{\mathcal{Y}}_{N} (h_{N_1}^L - h_{N_2}^L) d \varepsilon 
= 0.
\label{cond2}
\end{equation}
Equations (S16)-(S18) fully determine temperature dependence of the voltages $V_1$ and $V_2$ induced by the temperature gradient at the normal terminals. 

In general the above equations can only be solved numerically. However, at high enough temperatures $T_{1,2} \gg E_{\mathrm{Th}}$ a simple analytical solution becomes possible. As the term $h_{N_1}^T + h_{N_2}^T$ varies at the energy scale of order $T_{1,2}$ the main contribution to the integral in Eq. (S16) comes from energies $|\varepsilon| \sim T_{1,2}$ where the function $Q(\varepsilon)$ already reduces to a constant. Hence, in the leading order in $E_{\mathrm{Th}} / T_{1,2}$ Eq.~(S16) becomes essentially equivalent to the condition 
\begin{gather}
\int (h_{N_1}^T + h_{N_2}^T)
d \varepsilon = 0,
\end{gather}
which immediately yields $V_1=-V_2$.

Employing the same arguments we evaluate the first integral in Eq. (S18). Convergence of the second integral in this equation is controlled by the function $ G^{\mathcal{Y}}_{N}$ decaying at the scale of order Thouless energy. As a result, with a good accuracy we can expand $h_{N_1}^L - h_{N_2}^L$ up to the first nonvanishing order in $\varepsilon$ and rewrite Eq. (S18) as
\begin{equation}
\dfrac{\sigma \mathcal {A}_N}{L_N} (2 eV_1 - eV_2)
+
\dfrac{1}{2}
\left(
\dfrac{1}{T_1} - \dfrac{1}{T_2}
\right)
\int G^{\mathcal{Y}}_{N} \varepsilon d \varepsilon 
= 0.
\label{cond22}
\end{equation}
Since $V_1=-V_2$, we immediately get
\begin{gather}
e V_1 = - eV_2= - \dfrac{1}{8}
\left(
\dfrac{1}{T_1} - \dfrac{1}{T_2}
\right)
\dfrac{L_N}{\sigma \mathcal {A}_N} 
\int G^{\mathcal{Y}}_{N} \varepsilon d \varepsilon.
\label{V1V2A}
\end{gather}

\begin{widetext}
We can now relax the symmetry conditions for our X-junction and generalize the whole analysis to structures with normal wires of arbitrary lengths. After a simple algebra we derive the current conservation condition in the form
\begin{multline}
0
=
\int
\dfrac{(G_{S_1} + G_{S_2}) d\varepsilon}{
\det|\hat G_{S_1} + \hat G_{S_2} + \hat G_{N_1} + \hat G_{N_2}|}
\Bigl[
[G_{N_1}^T (G_{N_1}^L + G_{N_2}^L )+ G_{N_1}^{\mathcal{Y}} (G_{N_2}^{\mathcal{Y}} + G_{N_1}^{\mathcal{Y}})] h_{N_1}^T
+\\+
[G_{N_2}^T (G_{N_1}^L + G_{N_2}^L )+ G_{N_2}^{\mathcal{Y}} (G_{N_2}^{\mathcal{Y}} + G_{N_1}^{\mathcal{Y}})] h_{N_2}^T
-
(G_{N_1}^L  G_{N_2}^{\mathcal{Y}} - G_{N_2}^L G_{N_1}^{\mathcal{Y}}) (h_{N_1}^L - h_{N_2}^L)
\Bigr],
\label{cond1gen}
\end{multline}
cf. Eq. (S16). Equation (S18) can be generalized analogously, and we have
\begin{multline}
0 = 
\int
\dfrac{d \varepsilon}{
\det|\hat G_{S_1} + \hat G_{S_2} + \hat G_{N_1} + \hat G_{N_2}|}
\Biggl\{
[(G_{S_1} + G_{S_2}) + 2 G_{N_2}^T ] (G_{N_1}^L + G_{N_2}^L) G_{N_1}^T h_{N_1}^T
+\\+
[(G_{S_1} + G_{S_2}) (G_{N_1}^{\mathcal{Y}} - G_{N_2}^{\mathcal{Y}}) G_{N_1}^{\mathcal{Y}} 
+
2 G_{N_1}^T G_{N_2}^{\mathcal{Y}} G_{N_2}^{\mathcal{Y}} 
+
2 G_{N_2}^T G_{N_1}^{\mathcal{Y}} G_{N_1}^{\mathcal{Y}} ] h_{N_1}^T
-\\-
[(G_{S_1} + G_{S_2}) + 2 G_{N_1}^T ] (G_{N_1}^L + G_{N_2}^L) G_{N_2}^T h_{N_2}^T
-
[-(G_{S_1} + G_{S_2}) (G_{N_1}^{\mathcal{Y}} - G_{N_2}^{\mathcal{Y}}) G_{N_2}^{\mathcal{Y}} 
+
2 G_{N_2}^T G_{N_1}^{\mathcal{Y}} G_{N_1}^{\mathcal{Y}} 
+
2 G_{N_1}^T G_{N_2}^{\mathcal{Y}} G_{N_2}^{\mathcal{Y}} ] h_{N_2}^T
+\\+
[2 G_{N_2}^{\mathcal{Y}} (G_{N_1}^T G_{N_1}^L + G_{N_1}^{\mathcal{Y}} G_{N_1}^{\mathcal{Y}}) 
+
2 G_{N_1}^{\mathcal{Y}} (G_{N_2}^T G_{N_2}^L + G_{N_2}^{\mathcal{Y}} G_{N_2}^{\mathcal{Y}})
+
(G_{S_1} + G_{S_2})(G_{N_1}^L G_{N_2}^{\mathcal{Y}} + G_{N_2}^L G_{N_1}^{\mathcal{Y}})
] ( h_{N_1}^L - h_{N_2}^L )
\Biggr\},
\label{cond2gen}
\end{multline}
\end{widetext}
where we define
\begin{equation}
\hat G_{S_{1,2}} = 
\begin{pmatrix}
G_{S_{1,2}} & 0 \\
0 &0
\end{pmatrix}.
\end{equation}
At temperatures $T \gg E_{\mathrm{Th}}$ with a sufficient accuracy we can replace the coefficients  in front of $h_{N_{1,2}}^T$ in Eqs. (S22),~(S23) by their normal state values and employ the low energy expansion of $h_{N_{1,2}}^L$. Resolving these equations we get
\begin{equation}
e V_1=
\dfrac{1}{4}
\left(\dfrac{1}{T_1} - \dfrac{1}{T_2}\right)
\int
K_1(\varepsilon, \chi) \varepsilon d \varepsilon,
\label{V1gen}
\end{equation}
where
\begin{multline}
K_1(\varepsilon, \chi)=
\dfrac{1}{
\det|\hat G_{S_1} + \hat G_{S_2} + \hat G_{N_1} + \hat G_{N_2}|}
\times\\\times
\dfrac{1}{G_{N_1}^n }
\Biggl\{
G_{N_2}^{\mathcal{Y}} G_{N_1}^L 
\left[ 
G_{N_1}^n \dfrac{G_{S_1} + G_{S_2}}{G^n_{S_1} + G^n_{S_2}} - G_{N_1}^T 
\right]
-\\-
G_{N_1}^{\mathcal{Y}} G_{N_2}^L  
\left[
G_{S_1} + G_{S_2}
+
G_{N_1}^n \dfrac{G_{S_1} + G_{S_2}}{G^n_{S_1} + G^n_{S_2}}
+
G_{N_2}^T \right]
-\\-
G_{N_1}^{\mathcal{Y}} G_{N_2}^{\mathcal{Y}} ( G_{N_1}^{\mathcal{Y}} + G_{N_2}^{\mathcal{Y}} )
\Biggr\}.
\label{K1gen}
\end{multline}
The results for $V_2$ and $K_2(\varepsilon, \chi)$ are obtained from Eqs. (S25),~(S26) by interchanging the indices $1 \leftrightarrow 2$. 

Equations (S25),~(S26) determine a formally exact expression for the thermally induced voltage $V_1$ in the leading order in $1/T$. We observe that non-vanishing thermoelectric effect in our X-junction arises only provided the function $\mathcal{Y}$ differs from zero. In the limit $T_{1,2} \gg E_{\mathrm{Th}}$ we may replace the spectral conductances $G_{S_{1,2}}$ and $G_{N_{1,2}}^{T,L}$ by their normal state values and neglect higher orders terms in $\mathcal{Y}$. Then Eq. (S26) reduces to
\begin{align}
K_1(\varepsilon, \chi)=
-
r_{N_1}
\int_{L_{N_1}} \mathcal{Y} \dfrac{dx}{L_{N_1}}.
\end{align}
Substituting this result into Eq.~(S25) we immediately arrive at Eq.~(11). The function $\mathcal{Y}$ here can be evaluated both numerically and analytically with the aid of the high energy expansion for the anomalous Green function. The latter procedure allows to explicitly perform both integrals over $x$ and $\varepsilon$ in Eq.~(11) and arrive at Eq.~(12).

\subsection{III. Long-range Josephson effect}
The functions $j_{\varepsilon}$ and $W(\varepsilon)$ can be evaluated analytically at sufficiently high energies 
$|\varepsilon| \gg E_{\mathrm{Th}}$. We obtain 
\begin{multline}
j_{\varepsilon} = 
\dfrac{16\varkappa}{3 + 2\sqrt{2}}
k e^{-kL_S}
\times\\ \times(\cos k L_S - \sin k L_S) \sin \chi \sgn \varepsilon,
\end{multline}
and 
\begin{multline}
W(\varepsilon)
=
\dfrac{2}{3 + 2\sqrt{2}}
r_{N_1}  r_{N_2}
\dfrac{L_{N_1} - L_{N_2}}{L_{N_1}  L_{N_2}}
\dfrac{ \varkappa^2 }{k }
e^{-kL_S}
\times\\\times
\left[
2 + \cos k L_S - \sin k L_S
\right] \cos^2(\chi/2),
\end{multline}
where $k = \sqrt{|\varepsilon|/D}$. Combining the above results one can easily verify that the product $j_{\varepsilon} W(\varepsilon) \varepsilon$ remains regular if extrapolated to lower energies. Employing this extrapolation we evaluate the corresponding energy integral approximately as
\begin{multline}
\int
W(\varepsilon) j_{\varepsilon}
\varepsilon
d \varepsilon
\approx
- 
\dfrac{32}{(3 + 2\sqrt{2})^2}\dfrac{1101}{1250}
r_{N_1}  r_{N_2}
\times\\\times
\dfrac{L_{N_1} - L_{N_2}}{L_{N_1}  L_{N_2}}
E_{\mathrm{Th}}^2 \varkappa^3  
\sin \chi \cos^2(\chi/2),
\label{WapptoxA}
\end{multline}
and arrive at Eq.~(18).

\begin{figure}[hbt!]
\includegraphics[width=80mm]{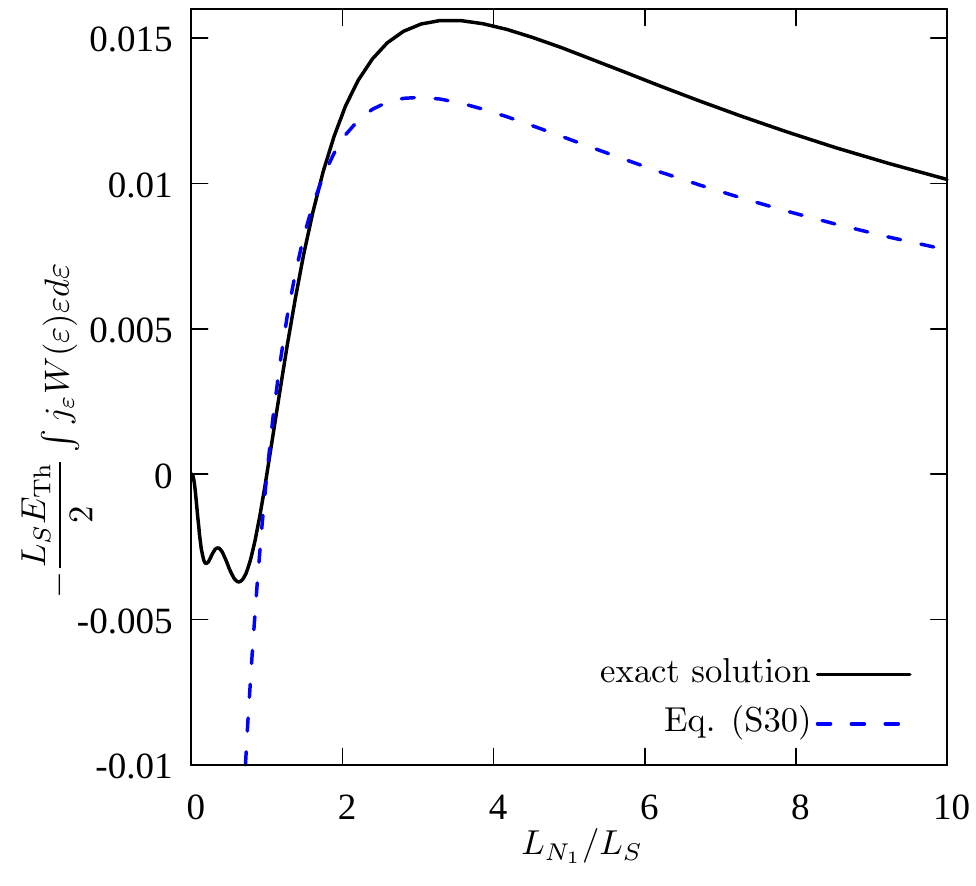}
\caption{Nonequilibrium contribution to the Josephson current as a function of $L_{N_1}$ for $\chi=\pi/2$, $L_{S_1}=L_{S_2}=L_S/2$, $L_{N_2}=L_S$ and  $\mathcal{A}_{S_1} = \mathcal{A}_{S_2} = \mathcal{A}_{N_1} = \mathcal{A}_{N_2}$.}
\label{W-LN2-fig}
\end{figure}

By comparing the above analytic result with the numerically exact one we can illustrate the accuracy of our simple approximation (S30). Figure S1 demonstrates that this approximation remains sufficiently accurate for relatively long normal wires with $L_{N_{1,2}} \gtrsim L_S$ and starts to fail for shorter values $L_{N_{1,2}}$.

\end{document}